\documentclass{easychair}

\usepackage{doc}

\usepackage{multirow}

\title{Frame Replication and Elimination for Reliability in Time-Sensitive Networks%
  }

\author{
    Peter Danielis\inst{1}
\and
    Helge Parzyjegla\inst{1}
\and
   Gero M\"uhl\inst{1}
\and 
   Eike Schweissguth\inst{2}
\and 
   Dirk Timmermann\inst{2}
}

\institute{
  Institute of Computer Science,
  University of Rostock,
  Rostock, Germany\\
  \email{\{peter.danielis;helge.parzyjegla;gero.muehl\}@uni-rostock.de}
\and
  Institute of Applied Microelectronics and Computer Engineering,\\
  University of Rostock,
  Rostock, Germany\\
  \email{\{eike.schweissguth;dirk.timmermann\}@uni-rostock.de}
 }

\authorrunning{Danielis, Parzyjegla, M\"uhl, Schweissguth, Timmermann}

\titlerunning{Frame Replication and Elimination for Reliability}

\begin{document} 

\maketitle

\begin{abstract}
In modern applications such as in the prospective smart factory, timely and fault-free communication is one of the main concerns. 
Communication failures may lead to huge economic losses. 
Moreover, they can even endanger human life. 
Therefore, the Time-Sensitive Networking (TSN) task group has introduced new standards for real-time capable Ethernet, which also include a fault tolerance mechanism called Frame Replication and Elimination for Reliability (FRER) as IEEE standard 802.1CB.
This standard introduces procedures and protocols for bridges and end stations in time-sensitive networks. 
It also provides mechanisms for the identification and duplication of frames to enable redundant transmissions.
In this paper, a simulation model is developed that implements the IEEE 802.1CB standard in OMNeT++.
In addition, as supplement to the standard we propose a reliability mechanism for establishing redundant paths and an error model to model transient and permanent errors.
As proof of concept, we evaluate the model with different topologies under various conditions.
\end{abstract}

%
%

\section{Introduction}
\label{sect:introduction}

For safety-critical applications, reliability is an important feature for any frame delivered from source to destination. 
A delay of a frame or its unsuccessful delivery can cause significant effects on the whole network's performance. 
For instance, if there is a fire in a building, the fire alarm system must deliver the frame to the concerned department on time to initiate rescue operations. 
Furthermore, in case of dealing with heavy machinery, a fault in any lifting machine must be reported timely since it can otherwise cause hazardous effects in the automation system. 
In such applications, if there is any fault in the link or bridge, the time required for retransmission of a lost frame over another path is intolerable. 
Therefore, the IEEE 802.1CB Time-Sensitive-Networking (TSN) standard, also refered to as Frame Replication and Elimination for Reliability (FRER), proposes to duplicate frames and to send them over multiple paths from sender (also refered to as talker) to receiver (also denoted as listener)~\cite{8091139}.
The standard extends the reliability feature of the IEEE 802.1Q standard~\cite{8403927}.
In case of a fault on one or more routes, the other routes can still deliver the frame to the destination.
In that way, the standards aims at increasing the probability of a frame delivery. 
The IEEE 802.1CB standard replicates a frame stream into one or more linked member streams, making the original stream a compound stream. 
An R-tag, a reserved field, and a sequence number that is unique for each frame are added to the frames. 
The frames of the original stream are replicated using the same sequence number, splitting the copies into the member streams and then rejoining those members stream at one or more points in the network. 
Duplicates are elimated based on the sequence number and the reconstituted stream are carried from those points.

In this paper, we present the implementation of the IEEE 802.1CB standard in OMNeT++ and evaluate it.
Further, some issues that are not addressed in the IEEE 802.1CB standard are considered. 
The standard only specifies how a frame is duplicated and how to do the elimination of those duplications. 
There is a need to specify which traffic needs to be duplicated and how many times to achieve the desired reliability. 
By means of an error model that is able to model transient and permanent errors, we determine the number of redundant paths to be established depending on the requested reliablity.
Also, we consider to establish a new path between sender and receiver if one of the paths for data transmission fails.

The remainder of this paper is organized as follows: In Section \ref{sect:related_work}, we compare our approach with the related work.
In Section~\ref{sect:supplements}, we describe our proposed supplements to the FRER standard. 
In Section \ref{sect:implementation_evaluation}, we provide an overview of the implementation of our proposed simulation model in OMNeT++ and evaluation results.
Finally, we conclude the paper in Section~\ref{sect:conclusions_future_work}, and present directions for future work.

\section{Related Work}
\label{sect:related_work}
In communication, reliability is an important feature for any frame to be delivered from source to destination. 
A delay of a frame or its unsuccessful delivery can cause significant effects on the whole network. 
To achieve such reliability, there are different protocols such as Parallel Redundancy Protocol (PRP)~\cite{kirrmann2007iec}, High Availability Redundancy Protocol (HSR)~\cite{kirrmann2011seamless}, and Media Redundancy Protocol (MRP)~\cite{giorgetti2012performance}.
These techniques have advantages over other protocols in providing reliable communication but there are other disadvantages when using those techniques.
In PRP, there is a need for a separate network to enable duplicated frame so additional infrastructure and thus costs are required. 
The disadvantage with HSR is that it can only be used for a ring topology.
For other topologies, HSR does not provide its full potential.
For that case, there is a need for a standard such as IEEE 802.1CB that provides reliable and timely communication, does not require additional infrastructure, and can be used for different topologies.
There are some works using the IEEE 802.1CB standard for different applications.
The authors in \cite{prinz2018end} use the standard for ensuring end-to-end redundancy between real-time Industry-4.0 components.
The configuration to send over redundant paths was integrated into the administration shell of real-time Industry-4.0 components as specific sub-model. 
The authors conduct performance measurements of the redundant transmission offset between the primary frame and its duplicate depending on the frame size, bandwidth, and number of network interfaces.
In \cite{pahlevan2018redundancy}, the FRER standard is used for the redundancy management of safety-critical applications using the Riverbed simulation framework.
To evaluate the results for safety-critical applications, a train network is used.
The model proves that the IEEE 802.1CB standard protects against transient error, offers bounded end-to-end delay, and can prevent frame loss.

\section{Supplements to the FRER Standard}
\label{sect:supplements}

Certain aspects are not addressed in the FRER standard and are therefore investiged in this paper. 
Some unaddressed issues which are not explained in the standard are as follows.
\begin{itemize}
\item Which traffic to replicate?
\item How many times to replicate the traffic?
\item What happens if a path fails permanently?
\end{itemize}

\subsection{Which Traffic to Replicate?\label{sect:which_traffic_replicate}}
The standard addresses how a frame is replicated and how replicates are eliminated.
However, the question arises if all the traffic should be replicated or only some priority classes. 
If all the traffic was replicated, an enormous burden would be put on the network and the network bandwidth is limited.
In case of a huge traffic load, the network might lose some important frames while the less important frames could consume all the network bandwidth.
The IEEE 802.1Q standard provides different classes of frames, from priority zero to priority seven~\cite{8403927}.
For instance, priority four, i.e., a video requires less than 100ms latency and jitter. 
Whenever a link of a route is disconnected, some time to establish a new route is required. 
This process may exceed the required latency.
Therefore, to address such a scenario, seamless redundancy is needed. 
Similarly, priority five, i.e., voice traffic demands less than 10ms latency, which requires a higher degree of redundancy. 
In this paper, we propose to replicate the frames with priority four and above, i.e, to use seamless redundancy for applications with real-time requirements.
\subsection{How Many Times to Replicate the Traffic?\label{sect:replicate_traffic}}

For a frame with a certain priority arrives, it has to be decided how many times the frame has to be replicated. 
The decision could be taken by a central SDN controller that is able to configure the network accordingly.
We propose to combine different methods to determine the number of duplications based on the frame's priority.
Firstly, we have to assign the reliability to different frame priorities, with which the frame should arrive at its destiniation in the presence of errors. 
The higher the priority, the higher is the reliability. 
We have chosen the reliability of different priorities according to Table~\ref{tab:frame_priorities}.
Please note that the chosen reliability values can be changed depending on application requirement.

\begin{table}[htp]
	\begin{centering}
		\begin{tabular}{l|r}
		\hline
		Frame priority & Reliability [\%]\\
		\hline
		4      &  99 \\
		5      &  99.9 \\
		6      &  99.99 \\
		7      &  99.999 \\
		\hline
		\end{tabular}
		\caption{Frame priorities and the frame's desired reliability to arrive at its destinination.}
		\label{tab:frame_priorities}
	\end{centering}
\end{table}

To match the assigned reliability of a frame with a path's probability to fail, the path failure probability needs to be calculated.
In the path, bridges have been attached with interconnecting links. 
For the calculation of bridge failure and link failure, the Norman Finn method is used~\cite{finn2013failure}. 
The probability of any bridge to fail ($T_f$) is assumed to ten thousand hours, and the meantime to repair any bridge ($T_r$) is assumed to twenty-four hours. 
The probability of frame loss due to bridge failure rate ($N_f$) is calculated, as shown in Equation~\ref{eq:bridge_failure_rate}.
\begin{equation}\label{eq:bridge_failure_rate}
N_f = \frac{T_r}{T_r + T_f} = \frac{24h}{24h + 1000h} = 0.0024
\end{equation}
Through this formula, the bridge failure rate is calculated as 0.0024. 
The link failure rate ($L_f$) is assumed as 0.0001~\cite{finn2013failure}.
After calculating the bridge and link failure rates, we use the method proposed by the authors in~\cite{xu2016failure} to calculate the probability of the serial path to fail ($P_{pf}$). 
According to this method, the probability of a serial path is calculated by summing up all bridge failures rate ($N_{fi}$) and link failure rates ($L_{fi}$), as shown in Equation~\ref{eq:failure_probability_serial_path}. 
In the equation, $n$ shows the total number of bridges in the path.
\begin{equation}\label{eq:failure_probability_serial_path}
P_{pf} = \sum_{i=1}^n N_{fi} + \sum_{i=1}^n L_{fi}
\end{equation}
%

After the failure-probability calculation of the serial path, we need to calculate the failure probability of a parallel path, which we can derive from the reliability of the path. 
To calculate the reliability, the formula proposed in the work~\cite{koren2020fault} is used as shown in Equation~\ref{eq:reliability_parallel_path}. 
The parallel path connected with $n$ modules requires all $n$ modules to fail to cause whole network failure. 
\begin{equation}\label{eq:reliability_parallel_path}
R_{p} = 1 - \prod_{i=1}^n (1-R_{i})
\end{equation}
Thereby, $R_p$ donates the path reliability, $R_i$ donates the reliability of the bridges and links, and $n$ donates the total number of bridges.
%
%
However, there are also scenarios, in which a combination of series and parallel paths is used~\cite{koren2020fault}.
The reliability of these systems can be calculated as expressed in Equation~\ref{eq:reliability_non_series_path}.
\begin{equation}\label{eq:reliability_non_series_path}
R_{sys} = 1 - \prod_{i=1}^n (1-R_{pi})
\end{equation}
In this equation, $R_{sys}$ is referred to as the reliability of the whole system, and $R_{pi}$ is referred to as the bridges' reliability. 
%
%
%
\begin{figure}[tb]
  \begin{centering}
    \includegraphics[width=0.8\textwidth]{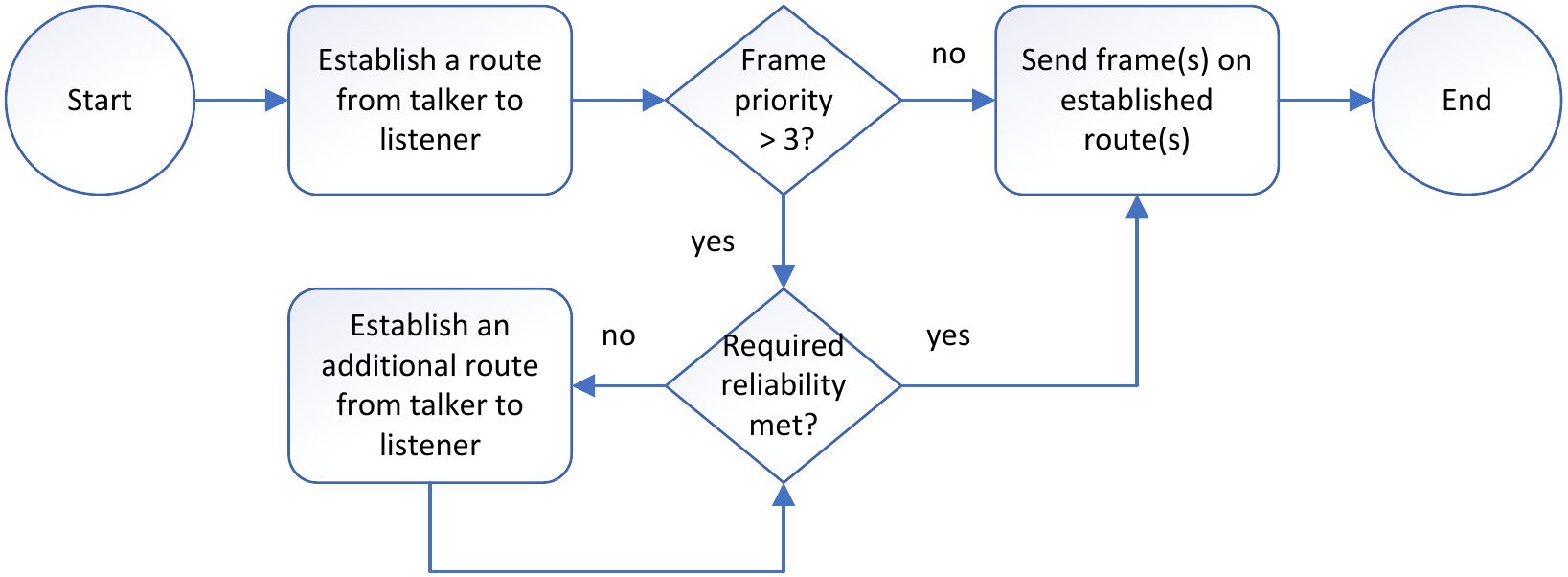}
  \end{centering}
  \caption{Algorithm to establish the required number of redundant routes depending on the desired reliability.}
  \label{fig:algorithm1}
\end{figure}
Considering the probability calculation of the system, there is a need to establish the required number of redundant routes in such a way that the probability that a frame arrives from talker to the listener is according to the desired reliability. 
Therefore, an algorithm is proposed to establish the required number of redundant routes. 
The algorithm (see Figure~\ref{fig:algorithm1}) that could be executed by a central SDN controller that is able to configure the network is explained in the below-stated steps.
\begin{enumerate}
    \item Establish a route selected from the available paths of the topology from talker to listener.
    \item The FRER functionality checks whether the priority of a frame is above 3 (cf. Table~\ref{tab:frame_priorities}).
    \item If the priority is three or less, the frame is sent on the first route and the algorithm terminates.
    \item If the priority is greater than three, it is checked whether the required probability is met.
    \item If the required probability is met, then the frame is sent on the established route(s).
    \item If not and additional routes are available, an additional route is established and it is checked again whether the required probability is met.
    \item If the required probability is achieved or no more additional routes are available, then the frame is sent on the established routes. Otherwise, step 4 is repeated until the probability is met.
\end{enumerate}
Now the problem arises of how to find the new route. 
For this purpose, a few conditions must be considered as follows.
\begin{itemize}
    \item If it is the first route, find the route with the highest reliability, which means the lowest failure rate.
    \item For further routes, find the route with the highest reliability among the non-selected paths.
    \item If two route shave the same reliability, then choose the one with the most independent bridges or links.
\end{itemize}
%
\subsection{What happens if a path fails permanently?\label{sect:permanent_error}}
\begin{figure}[tb]
  \begin{centering}
    \includegraphics[width=0.8\textwidth]{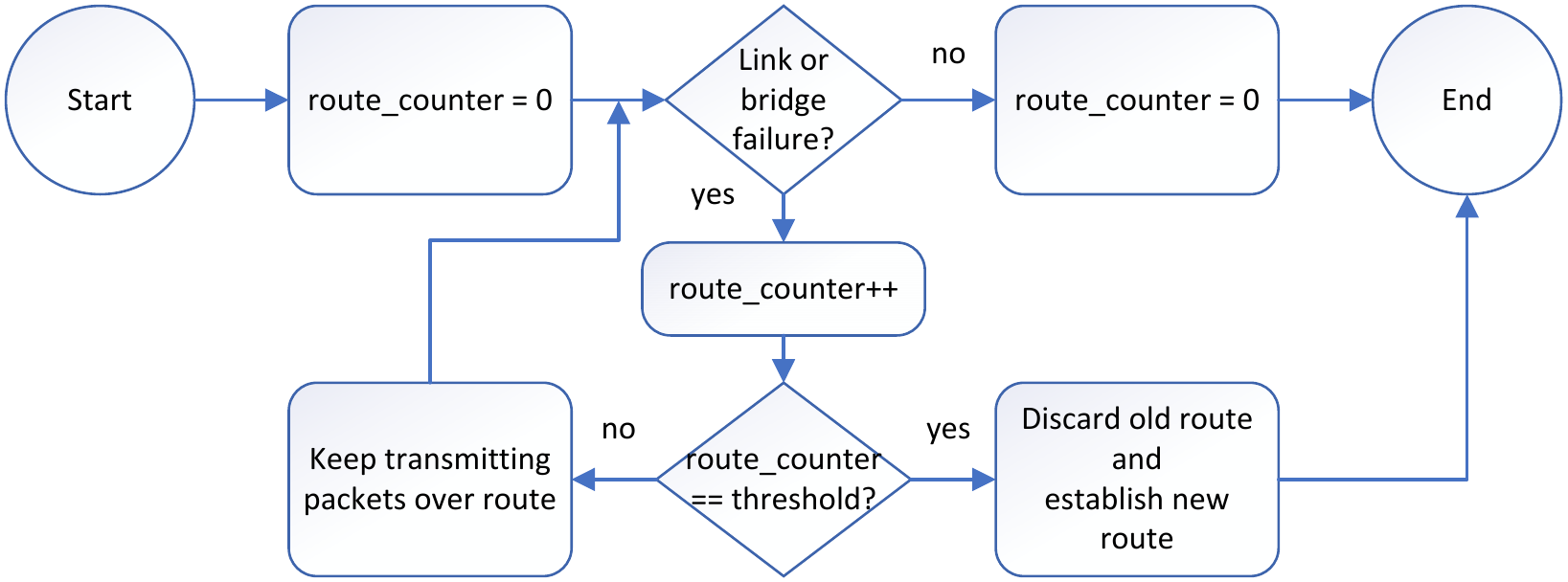}
  \end{centering}
  \caption{Algorithm to find and counter permanent route errors.}
  \label{fig:algorithm2}
\end{figure}

The third issue that is not addressed in the standard is what happens if a route fails permanently. 
There is a need to establish a new route while discarding the failing route.
In this paper, we consider two kinds of errors. 
The first one is the transient error. 
If this occurs, there is an occasional frame drop on a route. 
Either there is a fault on a link or an error in a bridge. 
The second kind of error is a permanent error. 
If this error happens either the link of the route is permanently broken or there is a faulty bridge.
This can cause a permanent fault on the route and hence the frame can no longer be sent on this path.
A frame duplication can counter a transient error. 
To counter it, firstly check the bridge and link reliability and then decide how many times a frame has to be duplicated. 
To deal with the second kind of error, in which a route fails permanently, an algorithm is proposed as apparent from Figure~\ref{fig:algorithm2}, which is described as follows.
\begin{itemize}
    \item If a link or bridge is unable to forward a frame, increase a counter for that route. A central SDN controller used for configuring the network could keep track of the counter.
    \item Check if the counter is equal to a treshold value, e.g., three. Please note that the threshold value can be set to an arbitrary value depending on  application requirements. 
    \item If the counter is not three, then keep sending a frame over that route.
    \item If the counter is equal to three, discard the faulty route and choose a new route from the remaining paths of the topology if available. A central SDN controller could reconfigure the network in such case.
    \item To choose a new route, a previously described algorithm is used, as explained in the previous Section~\ref{sect:replicate_traffic}.
    \item If the counter is less than three and the route has recovered from the transient error so that the packet is successfully transmitted over the route again, then reset the counter.
\end{itemize}
According to the proposed algorithm, there is a need to precompute additional routes other than the actively used routes, so in case of permanent failure, there is an extra route for the network to utilize it.
%
\section{Implementation and Evaluation}
\label{sect:implementation_evaluation}
This section outlines the OMNeT++ implementation of the FRER standard and the supplements as proposed in Section~\ref{sect:supplements}.
Further, the simulation setup is sketched and results are given and evaluated obtained from simulation runs in OMNeT++.
\subsection{Implementation}
\begin{figure}[tb]
  \begin{centering}
    \includegraphics[width=0.3\textwidth]{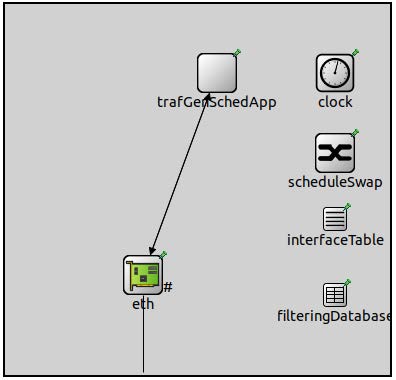}
    \includegraphics[width=0.35\textwidth]{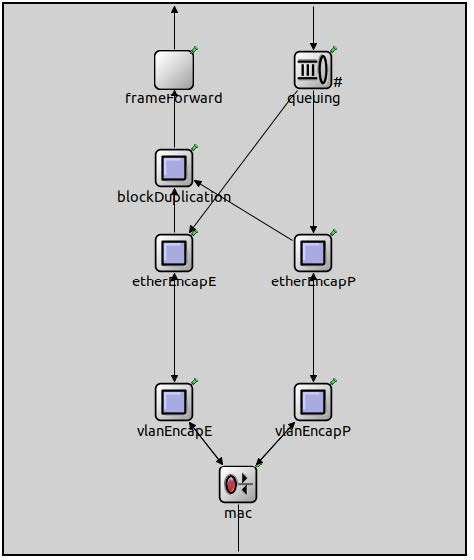}
  \end{centering}
  \caption{OMNeT++ functionality of a talker (left) and a bridge (right).}
  \label{fig:OMNeT++_functionality}
\end{figure}

The FRER functionality is added to a talker and a bridge as implemented in the NeSTiNG project~\cite{falk2019nesting}. 
The talker function is expressed as the \emph{trafGenSchedApp} module in Figure~\ref{fig:OMNeT++_functionality} (left). 
The \emph{trafGenSchedApp} module has been used previously to generate the frame in the IEEE 802.1Q frame format~\cite{falk2019nesting}. 
Now, the module is extended and is also used to generate sequence numbers and to add an R-tag into the frame. 
After the frame has been generated by adding the R-tag into the frame, it is passed to the \emph{eth} module, where it passes the frame to the MAC layer.
At the talker, the clock function is used to add the time to the frame, and the \emph{filteringDatabase} function is used to load the routing file (.xml).
In the bridge, there are multiple functions, as explained in the FRER standard. Numerous functions were already present in the NeSTiNG model and some are added to the bridge. 
The complete structure of the bridge, as implemented in OMNeT++, is shown in Figure~\ref{fig:OMNeT++_functionality} (right). 
All other functions except \emph{blockDuplication} already existed in the NeSTiNG switch.
This module is extended with functionality to understand the FRER-specific information so the bridge can decide how to handle the FRER frames wherever necessary according to the FRER functionality. 
The switch module, as implemented, is described as follows.
\begin{itemize}
    \item \emph{queuing} module: It is used to maintain the frame's queue to decide which frame will be forwarded next.
    \item \emph{etherEncap}: This function checks the basic parameter of the IEEE 802.1Q parameters. There are two functions of \emph{etherEncap}, both doing the same task.
    \item \emph{vlanEncap}: This function checks the basic frame parameters of FRER standard.
    \item \emph{blockDuplication}: This function checks for duplicated frames and discards it if a duplicated frame was received.
    \item \emph{frameForward}: This function receives the frame from the lower layer and forwards it to the \emph{processingDelay} function of the NeSTiNg bridge (cf.~\cite{falk2019nesting}).
    \item \emph{Mac}: This function takes the frame from the upper layer and passes it to the next link.
\end{itemize}
%
\subsection{Simulation Setup}
\begin{figure}[tb]
  \begin{centering}
    \includegraphics[width=0.32\textwidth]{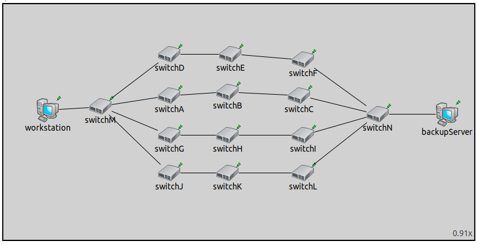}
    \includegraphics[width=0.32\textwidth]{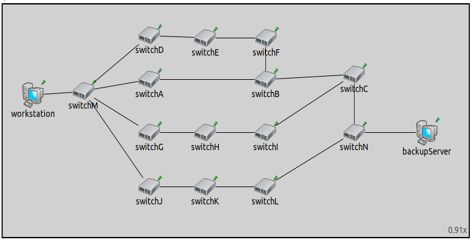}
    \includegraphics[width=0.32\textwidth]{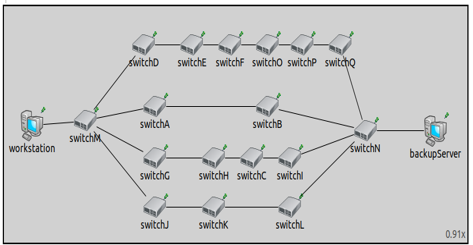}
  \end{centering}
  \caption{Topologies evaluated in OMNeT++ (left: topology 1, middle: topology 2, right: topology 3).}
  \label{fig:OMNeT++_topologies}
\end{figure}

As proof of concept, we evalute results from three different topologies as apparrent from Figure~\ref{fig:OMNeT++_topologies} (left: topology 1 with four parallel paths, middle: topology 2 with interconnecting links between four parallel paths, right: topology 3 with different numbers of bridges per one of the four parallel paths).
The paths are enumerated with 1 to 4 from top to down.
Seven test cases without errors, with a transient fault, and with a permament fault are evaluated, cf. Table~\ref{tab:test_cases}.
\begin{table}[htp]
	\begin{centering}
		\begin{tabular}{c|c|c|c|c}
		\hline
		Test case & Topology & Error & Frame priority & Established routes over path(s)\\
		\hline
		1 & 1 & None & 3 & Path 1\\
		2 & 1 & None & 4, 5 & Path 1 and 2\\
		3 & 1 & None & 6, 7 & Paths 1, 2, and 3\\
		4 & 2 & None & 5 & Path 2 and 4\\
		5 & 3 & None & 5 & Path 2 and 4\\
		6 & 3 & Transient & 5 & Path 2 and 4\\
		\multirow{2}{*}{7} & \multirow{2}{*}{3} & \multirow{2}{*}{Permanent} & \multirow{2}{*}{5} & Initially: path 2 and 4\\
		 &  & & & After error: path 3 and 4\\
		\hline
		\end{tabular}
		\caption{Description of test cases.}
		\label{tab:test_cases}
	\end{centering}
\end{table}
During each test case, 10 frames are sent every 30$\mu$s with different priorities (cf. Table~\ref{tab:frame_priorities}).
Depending on the selected priorities, frames are sent on only one route or redundantly on several routes selected from the available paths of the topologies.
Please note that the chosen topologies and test cases are solely examples to demonstrate the FRER functionality in OMNeT++.

%
\subsection{Evaluation}
The first three test cases are evaluated on topology 1 and no errors occur while the frames' priorites increase from the first to the third case.
Hence, one or more routes are needed to ensure the desired reliability.
In the first test case, for frames with priority 3 no redudant routes are needed according to Section~\ref{sect:which_traffic_replicate} so frames are solely sent on path 1 with a delay of 43.98$\mu$s.
In the second test case, frames with priority 4 and 5 are sent.
Following the statements in Section~\ref{sect:replicate_traffic}, two redundant routes established on path 1 and 2 are needed on topology~1.
The end-to-end delay of the frames is 43.98$\mu$s like in test case 1 since all paths contain five bridges and six links.
In test case 3, frames with priority 6 and 7 are sent on three routes in parallel (path 1, 2, and 3) with the same end-to-end-delay as in the previous test cases.

The fourth test case is evaluated on topology 2 with no errors while the priorities of the frames is set to 5 so that frames have to be sent on two routes in parallel.
We have selected the two routes established on path 2 and 4, which are equal in length, i.e., they consist of five bridges and six links each.
This again leads to an end-to-end delay of 43.98$\mu$s. 

In the fifth test case, frames with priority 5 are sent on two routes of topology 3 (path 2 and 4).
The route on path 2 is shorter than the other on path 4 and comprises four bridges and five links which leads a shorter end-to-end latency of 35.82$\mu$s as compared to the second routes with five bridges and six links.
Since no errors occur the frame traveling on the shorter route arrives before its duplicate arrives on the longer route and the duplicate is therefore discarded.

In the sixth test case, frames with priority 5 are sent on two routes in parallel on topology 3.
The routes are the same as those selected in test case 5 (path 2 and 4).
The test case incorporates a transient error in such a way that every third frame is dropped on the shorter route.
As expected, we observe that every third frame traveling on the longer route arrives at the listener later (43.98$\mu$s) since the frame on the shorter route is dropped.
However, the frame arrives since we have established two redundant routes.

In the seventh test case, frames with priority five are initially sent on the two routes established on path 2 and 4 on topology 3.
After the fifth frames has been sent, the route on path 2 fails permanently and frames are solely delivered on the route on path 4.
Since the route on path 4 is longer than the one on path 2, the end-to-end delay increases from 35.82$\mu$s to 43.98$\mu$s permanently after the fifth sent frame.
After the frame has failed to be sent three times (according to Section~\ref{sect:permanent_error}) on path 2, the path 2 is assumed to be broken permanently and path 3 is selected instead according to Section~\ref{sect:replicate_traffic}.
\section{Conclusions and Future Work}
\label{sect:conclusions_future_work}
In this paper, we proposed a simulation model for the IEEE 802.1CB standard for reliability in time-sensitive networks. 
The standard leaves some issues unaddressed which we considered in our model. Firstly, we derive from a frame's priority which frame should be duplicated. 
Secondly, we developed a model to address the question how many times a frame should be duplicated depending on its priority.
Thirdly, we proposed a mechanism to select a new route on which a frame is sent if there is a permanent failure in any of the routes. 
To evaluate the reliability of the network and failure model, different topologies are tested with different conditions.
We enforce different errors to check the model behavior in the presence of transient and permanent errors.
The simulation results show that that the model works as expected and protects against transient and permanent errors.
In our model, we used predefined paths and configurations. 
As part of future work, the model will be configured at runtime.
\section{Acknowledgments}
We would like to thank the German Research Foundation (DFG) (project "Real-Time Publish/Subscribe for the Industrial Internet of Things", GZ: TI 254) for their financial support.

\label{sect:bib}
\bibliographystyle{plain}
\bibliography{./bib/OSC_FRER}


\end{document}